\documentclass[twocolumn,trackchanges]{aastex631}
\usepackage{hyperref}
\usepackage{amsmath}
\usepackage{amssymb}
\usepackage{graphicx}
\usepackage{color}
\usepackage{subfigure}
\usepackage{bm}
\usepackage{textcomp,gensymb}
\usepackage{bigints}
\usepackage{aas_macros}
\usepackage{physics}
\usepackage{chemformula}

\hypersetup{linkcolor=blue,citecolor=blue,filecolor=blue,urlcolor=blue}
\submitjournal{ApJ}

\begin{document}

\title{Constraining fundamental parameters in modified gravity using {\it Gaia}-DR2 massive white dwarf observation}
\shorttitle{Fundamental constants in $f(R)$ gravity}
\shortauthors{Kalita \& Uniyal}

\correspondingauthor{Surajit Kalita}
%\email{greg.schwarz@aas.org, gus.muench@aas.org}

\author[0000-0002-3818-6037]{Surajit Kalita}
\affiliation{High Energy Physics, Cosmology and Astrophysics Theory (HEPCAT) Group, Department of Mathematics and Applied Mathematics, University of Cape Town, Cape Town 7700, South Africa}
\email{surajit.kalita@uct.ac.za}

\author[0000-0001-8213-646X]{Akhil Uniyal}
\affiliation{Department of Physics, Indian Institute of Technology, Guwahati 781039, India}
\email{akhil\_uniyal@iitg.ac.in}



\begin{abstract}
Various experiments and observations have led researchers to suggest different bounds on fundamental constants like the fine-structure constant and the proton-to-electron mass ratio. These bounds differ mostly due to the energy scale of the systems where the experiments are performed. In this article, we obtain bounds on these parameters in the modified gravity context using the {\it Gaia}-DR2 massive white dwarf data and show that the bounds alter as the gravity theory changes. This exploration not only indicates strong support for non-negligible influences of modified gravity in astrophysical scenarios in high-density regimes but also reveals that the bounds on fundamental parameters can be much stronger under alternate gravity theories.
\end{abstract}



\section{Introduction}
The success of general relativity (GR) is very prominent in a number of experiments~\citep{carroll_2019} though it has some limitations due to its viability primarily at the low-energy limit like the Newtonian theory is valid in weak-gravity regimes. One of the main shortfalls comes from the cosmological framework such as the simultaneous explanation of the accelerated expansion of the universe and the inflationary cosmology~\citep{SupernovaSearchTeam:1998fmf, SupernovaCosmologyProject:1998vns}. Moreover, it also predicts singularity at a small scale and has theoretical discrepancy that it is not a renormalizable theory~\citep{tHooft:1974toh}. Therefore, testing the gravity theories in the strong-field limit is still an open problem~\citep{Berti:2004bd}. One of the popular alternatives to GR is $f(R)$ gravity containing higher-order terms of scalar curvature $R$ in the Einstein-Hilbert action~\citep{Sotiriou:2008rp}.

In $f(R)$ gravity, there are two ways to obtain the modified field equations using the variation principle. First is simply varying the action with respect to the metric resulting in the fourth-order field equations and the other is Palatini formalism where the action is varied with respect to the metric as well as the affine connection when they are not related beforehand, which results in the second-order field equations~\citep{Fay:2007gg}. These two formalisms are equivalent in the case of the GR where $f(R)=R$. In recent times, people have studied Palatini $f(R)$ gravity in more detail because of the simplicity to handle second-order equations over fourth-order equations along with the fact that it is free from the ghost-like instabilities which usually appear in the metric formulation of the theories~\citep{Olmo:2012yv, 2014EPJC...74.2924L}. Some such works are in the cosmological framework~\citep{Sotiriou:2005hu} and others are in the astrophysical systems such as white dwarfs~(WDs)~\citep{2022PhRvD.105b4028S}, neutron stars~(NSs)~\citep{TeppaPannia:2016vsb, Herzog:2021wpj}, stellar~\citep{Olmo:2019qsj}, and substellar~\citep{Benito:2021ywe} objects. In this work, we use the Palatini $f(R)$ gravity formalism.

Moreover, the final evolution stage of a main sequence star with mass lying below $(10\pm2)\,\rm M_\odot$ is a WD~\citep{1996cost.book.....G,2018MNRAS.480.1547L}. Chandrasekhar showed that the mass of a non-rotating non-magnetized carbon-oxygen WD cannot exceed $1.44\,\textup{M}_\odot$~\citep{1935MNRAS..95..207C}, which is popularly known as the Chandrasekhar mass-limit. Now, if a WD accumulates mass more than this limit, it explodes and releases a tremendous amount of energy in the form of a type Ia supernova (SN\,Ia). However, the discoveries of several peculiar over- and under-luminous SNe\,Ia suggest that they originated from the super- and sub-Chandrasekhar mass WDs respectively~\citep{1992AJ....104.1543F,1997MNRAS.284..151M,2006Natur.443..308H,2010ApJ...713.1073S}. This indicates that the Chandrasekhar mass-limit is sacrosanct but may not be unique, and researchers have used modified gravity to explain super- and sub-Chandrasekhar mass-limits simultaneously. Over the last decade, different modified gravity theories including $f(R)$ gravity have been proposed to obtain these limits both in relativistic and non-relativistic regimes~\citep{2018JCAP...09..007K,2021ApJ...909...65K,2022PhRvD.105b4028S,2022PhLB..82736942K,2022PhRvD.106l4010A}.

The debate on the existence of a valid alternative theory of the GR is still going on and in that order, Palatini $f(R)$ gravity has been looked into a great detail starting from the formulation of this theory to the validation in the stellar structure physics. Earlier it was suspected that this theory suffers from the curvature singularity in the stellar structure models~\citep{Barausse:2007pn, Barausse:2007ys, 2008EAS....30..189B}, which makes a perception of ruling out the Palatini $f(R)$ gravity. However, this statement was not strong enough since it depends on the type of compact object such as whether it is a NS or a WD, and thus on the equation of state (EoS)~\citep{2007PhRvD..76b4020K,Olmo:2008pv}. In fact, it was argued that the model $f(R)=R \pm \gamma R^2$ with $\gamma$ being the modified gravity parameter, does not suffer any curvature singularity when $\gamma$ is of the order of the Planck length squared, implying that a single electron's presence in the Universe is sufficient to handle all such stellar singularities. However, for a larger length scale of $\gamma$, the singularity problem may still hold. Therefore, one can use the compact stars to put the specific bounds on the model parameter $\gamma$ but it certainly does not rule out all Palatini $f(R)$ gravity~\citep{Olmo:2008pv}. This further motivates researchers to look more into the validity of this theory and its use in stellar models. In order to do so, the analysis done by~\cite{Barausse:2007pn, Barausse:2007ys, 2008EAS....30..189B} can be improved by using the properties of stellar surfaces in polytropic EoS models with specific junction conditions~\citep{Olmo:2020fri} to match the stellar and vacuum solutions at the surface of the star with the refined mathematical approach based on the tensorial distribution. In fact, they showed that the trace of the stress–energy tensor in the bulk must be continuous over the matching hypersurface, although its normal derivative does not have to be. The previous works did not appropriately consider these dynamics in the junction condition; thereby facing the singularity problem. Furthermore, it has been shown in the work that NSs and WDs both can be safely modeled within the Palatini $f(R)$ gravity framework. Therefore, because of the successful counterarguments to the challenges available in the literature (see reviews by~\cite{Olmo:2011uz} and~\cite{Olmo:2019flu} for more details), we consider the Palatini $f(R)$ gravity with $f(R)=R+\gamma R^2$ to constrain the fine-structure constant $\alpha=e^2/(\hbar c)$ and proton-to-electron mass ratio $\mu=m_\text{p}/m_\text{e}$, where $e$ is the charge of an electron, $h$ is the Planck constant, $\hbar=h/2\pi$, $c$ is the speed of light, $m_\text{p}$ is the mass of a proton, and $m_\text{e}$ is the mass of an electron.

Cosmological tests show that the bound $\Delta \mu /\mu \le 10^{-5}$ at a redshift $2 \le z \le 3$ by using the \ch{H2} absorption systems~\citep{king2011new,Le:2019ijj}. Further, the analysis of \ch{NH3} from quasar PKS\,$1830-211$ spectrum at $z=0.89$ provide a better bound $\Delta \mu /\mu = (0.0 \pm 1.0) \times 10^{-7}$~\citep{2014PhRvL.113l3002B}. However, this technique is valid only on the low-redshift systems with $z \le 1.0$ for \ch{NH3} and \ch{CH3OH}~\citep{Rahmani:2012ze}. Having a higher redshift such as $z \approx 3.17$ towards J$1337+3152$, the bound becomes $\Delta \mu /\mu = (-1.7 \pm 1.7) \times 10^{-6}$ and $\Delta \mu /\mu = (0.0 \pm 1.5) \times 10^{-6}$ at $z_\text{abs} \approx 1.3$ with relating to the four $21$ cm absorption systems~\citep{Srianand:2010un}. There are other bounds obtained from different cosmological data, which can be found in~\cite{2015PhRvC..92a4319D,2015Ap&SS.357....4K,2017RPPh...80l6902M,2017PhRvC..96d5802M,2018MNRAS.474.1850H}. Moreover, researchers have also used WD data to constrain $\alpha$ and $\mu$. Studying the H$_2$ transition from the spectra of WDs GD\,133 and G$29-38$, the proposed bounds are $\Delta \mu /\mu = (-2.7 \pm 4.9) \times 10^{-5}$ and $\Delta \mu /\mu = (-5.8 \pm 4.1) \times 10^{-5}$ respectively~\citep{2014PhRvL.113l3002B}. Similarly, using Fe\,V data from another WD G191$-$B2B, the bound is $\Delta \alpha / \alpha = (6.36\pm2.27)\times10^{-5}$~\citep{2021MNRAS.500.1466H}. Further, in a recent work, the authors used a simulated catalog of 100 WDs mass--radius pair in the mass range $0.3\,\textup{M}_\odot < M < 1.2\,\textup{M}_\odot$ and found $\Delta \alpha / \alpha = (2.7 \pm 9.1) \times 10^{-5}$~\citep{2017PhRvD..96h3012M}. Note that this bound is the outcome of the simulated WD data, not an actual one. We show later that they are far from realistic data and hence this result is not trustworthy.

All these previously reported bounds primarily depend on the environment such as the redshift; thereby the energy scale of the system where the experiment is performed. This work, for the first time, shows possible variations of these fundamental parameters in a modified gravity inspired astrophysical system. In other words, we show that the bounds can change even though the energy scale of the system does not change. We use WD mass and radius data from {\it Gaia}-DR2 observation and constrain $\alpha$ and $\mu$ in $f(R)$ gravity. To do so, in Sec.~\ref{Sec:2}, we establish the hydrostatic balance equations for Palatini $f(R)$ gravity, which we revise to incorporate any deviation on $\alpha$ and $\mu$. We then compare the masses and radii of the reported WDs from the {\it Gaia}-DR2 observation with the theoretical mass--radius curves for $R+\gamma R^2$ model to constrain $\alpha$ and $\mu$ for different $\gamma$ in Sec.~\ref{Sec:3}. Finally, we discuss our results and put our concluding remarks in Sec.~\ref{Sec:4}.

%%%%%%%%%%%%%%%%%%%%%%%%%%%%%%%%%%%%%%%%%%%%%%%%%%%%%%%%%%%%%%%%%%%%%%%%%%%%%%%%%%%%%%%%%%%%%%%%%%%%%%%%%%%%%%%%%%%%%
\section{Palatini $f(R)$ gravity and stellar structure of white dwarfs}\label{Sec:2}

The action in the $f(R)$ gravity for a metric $g_{\mu \nu}$ is given by~\citep{Sotiriou:2008rp}
\begin{equation} \label{e2.1}
    \mathcal{S}[g,\Gamma,\psi]=\frac{1}{2\kappa} \int \sqrt{-g} f(R(g,\Gamma))\dd[4]{x} + \mathcal{S}_\text{m}[g,\psi],
\end{equation}
where $\kappa=8\pi G/c^4$, $g=\det(g_{\mu \nu})$, $\Gamma$ is the Levi-Civita connection, $G$ is Newton’s gravitational constant, and $\mathcal{S}_\text{m}$ is the matter action which depends on $g_{\mu\nu}$ and the matter field $\psi$. Now, in Palatini formalism, we vary this action first with respect to $g_{\mu \nu}$ and then with respect to $\Gamma$ assuming they are independent of each other to respectively obtain the following modified field equations~\citep{Sotiriou:2008rp}
\begin{align} \label{e2.21}
    f'(R)R_{\mu \nu}-\frac{1}{2}f(R)g_{\mu \nu}&=\kappa T_{\mu \nu},\\
    \nabla_\lambda \left(\sqrt{-g}f'(R)g^{\mu \nu}\right)&=0, \label{e2.22}
\end{align}
where $f'(R) = \dv*{f(R)}{R}$, $\nabla_\lambda$ is covariant derivative with respect to $\lambda$, and $T_{\mu \nu}$ is the energy-momentum tensor. Let us define a metric tensor $\Bar{g}_{\mu \nu}=f'(R)g_{\mu \nu}$ with $\Bar{g}=\det(\Bar{g}_{\mu\nu})$, such that Eq.~\eqref{e2.22} can be recast as
\begin{equation} \label{e2.3}
    \nabla_\lambda(\sqrt{-\Bar{g}}\Bar{g}^{\mu \nu})=0.
\end{equation}
Now, assuming the matter inside the WD is non-magnetized and behaves as a perfect fluid, the energy-momentum tensor can be written as
\begin{equation} \label{e2.4}
    T^{\mu \nu}=(\rho c^2+P)u^\mu u^\nu + P g^{\mu \nu},
\end{equation}
where $P$ and $\rho$ are the pressure and matter density of the fluid respectively. In this work, we consider the functional form of $f(R)$ as
\begin{equation} \label{e2.5}
    f(R)=R+\gamma R^2.
\end{equation}

Because WDs are big in size, one can consider the Newtonian treatment to understand their structures. Thus we can safely assume $\rho c^2\gg P$. Now, in the weak-gravity limit, expanding the metric tensors as $g_{\mu \nu}=\eta_{\mu \nu}+h_{\mu \nu}$ and $\Bar{g}_{\mu \nu}=\eta_{\mu \nu}+\Bar{h}_{\mu \nu}$, such that $|h_{\mu \nu}|, |\Bar{h}_{\mu\nu}|\ll| \eta_{\mu \nu}|$ and denoting $h_{00}=-2 \Phi /c^2$ with $\Phi$ being the gravitational potential for $f(R)=R+\gamma R^2$, one can obtain the following modified Poisson equation~\citep{2020PhRvD.101f4050T}
\begin{equation}\label{e2.6}
    \nabla^2 \Phi \approx 4\pi G (\rho - 2\gamma \nabla^2 \rho).
\end{equation}
This translates into obtaining the hydrostatic balance equations for a non-rotating spherically symmetric WD, given by~\citep{2023PhRvD.107d4072K}
\begin{align}\label{e2.8}
    \dv{P}{r} &= -\frac{Gm\rho}{r^2}+8\pi G\gamma \rho \dv{\rho}{r},\\
    \dv{m}{r} &= 4\pi r^2 \rho,\label{e2.9}
\end{align}
where $m$ is the mass accumulated inside a radius $r$. Note that for $\gamma=0$, we recover the well-known Newtonian hydrostatic balance equations. Note that although these equations are valid for Palatini $f(R)$ gravity, similar equations can also be obtained in general scalar-tensor-vector gravity or fourth-order gravity theories~\citep{2016PhRvL.116o1103J,2017JCAP...10..004B}. Thus the following calculations and the corresponding results are robust and more general in terms of modified gravity models. We further consider the matter inside a WD follows the Chandrasekhar EoS, given by~\cite{1935MNRAS..95..207C} as
\begin{equation}\label{e2.10}
\begin{split}
    P &= \frac{\pi m_\text{e}^4 c^5}{3 h^3}\left[x\left(2x^2-3\right)\sqrt{x^2+1}+3\sinh^{-1}x\right],\\
    \rho &= \frac{8\pi \mu_\text{e} m_\text{p}(m_\text{e}c)^3}{3h^3}x^3,
\end{split}
\end{equation}
where $x=p_\text{F}/m_\text{e} c$ with $p_\text{F}$ being the Fermi momentum, and $\mu_\text{e}$ the mean molecular weight per electron. Because we are only interested in obtaining the masses and radii of the WDs, which are predominantly outcomes of the core matter, we use this zero-temperature EoS and neglect any contribution from the ideal gas whose effects are prominent near the surface.

As we are interested in finding out the bounds on $\alpha$ and $\mu$ in $R+\gamma R^2$ gravity, let us define the dimensionless couplings $\alpha_\text{p}=Gm_\text{p}/\hbar c$ and $\alpha_\text{e}=Gm_\text{e}/\hbar c$. Considering that the particle mass and the QCD scale can vary but the Planck mass is fixed, we obtain the following equations for the uncertainties in electron and proton masses~\citep{Coc:2006sx}
\begin{align}\label{e2.11}
    \frac{\Delta \alpha_\text{e}}{\alpha_\text{e}}=2\frac{\Delta m_\text{e}}{m_\text{e}} &= \left(1+\mathsf{S}\right)\frac{\Delta \alpha}{\alpha},\\
    \label{e2.12}
    \frac{\Delta \alpha_\text{p}}{\alpha_\text{p}}=2\frac{\Delta m_\text{p}}{m_\text{p}} &= \left[\frac{8}{5}\mathsf{R} +\frac{2}{5} \left(1+\mathsf{S}\right)\right]\frac{\Delta \alpha}{\alpha},
\end{align}
where $\mathsf{R}$ and $\mathsf{S}$ are dimensionless phenomenological parameters. Thus the uncertainty in $\mu$ is given by
\begin{equation}\label{e2.13}
    \frac{\Delta \mu}{\mu} = \left[\frac{4}{5}\mathsf{R}-\frac{3}{10}\left(1+\mathsf{S}\right)\right]\frac{\Delta \alpha}{\alpha}.
\end{equation}
Note that the values of $\mathsf{R}$ and $\mathsf{S}$ are model-dependent such that their absolute values can vary from unity to several hundred. Using the WMAP data, it was suggested $\mathsf{R} \approx 36$ and $\mathsf{S} \approx 160$~\citep{Coc:2006sx}, whereas the dilaton-type model whose variations of fundamental couplings have been discussed by~\cite{nakashima2010constraining} suggests that $\mathsf{R} \approx 109$ and $\mathsf{S} \approx 0$. On the other hand,~\cite{2014MmSAI..85..113M} used astrophysical observations of a radio source PKS\,1413$+$135 to provide $\mathsf{R}=278\pm24$ and $\mathsf{S}=742\pm65$. Now, substituting Eq.~\eqref{e2.10} in Eqs.~\eqref{e2.8} and~\eqref{e2.9}, along with rewriting them in terms of the $\alpha_\text{p}$ and $\alpha_\text{e}$, we obtain
\begin{align}\label{e2.14}
    \dv{x}{r} &= -\frac{K_1 \frac{m}{r^2} \frac{\sqrt{1+x^2}}{x}}{1-K_3 \gamma x \sqrt{1+x^2}},\\
    \dv{m}{r} &= K_2 r^2 x^3,
\end{align}
where
\begin{equation}\label{e2.16}
    \begin{split}
        K_1 &= \frac{G \mu_\text{e}}{c^2} \sqrt{\frac{\alpha_\text{p}}{\alpha_\text{e}}},\\
        K_2 &= \frac{8 \mu_\text{e} c^5}{3hG^2} \sqrt{\alpha_\text{p}\alpha_\text{e}^3},\\
        K_3 &= \frac{16 \mu_\text{e}^2c^3}{Gh} \alpha_\text{p}\alpha_\text{e},
    \end{split}
\end{equation}
Because uncertainties are only in the quantities $\alpha_\text{p}$ and $\alpha_\text{e}$, they propagate to $K_1$, $K_2$, and $K_3$. Thus including these corrections, refined forms of Eqs.~\eqref{e2.10} and~\eqref{e2.11} are given by
\begin{align}\label{e2.17}
    \dv{x}{r} &= -\frac{K_1\left(1+\beta_1\right) \frac{m}{r^2} \frac{\sqrt{1+x^2}}{x}}{1-K_3\left(1+\beta_3\right) \gamma x \sqrt{1+x^2}},\\
    \dv{m}{r} &= K_2\left(1+\beta_2\right) r^2 x^3,\label{e2.18}
\end{align}
where
\begin{equation}\label{e2.19}
    \begin{split}
        \beta_1&=\left[ \frac{4}{5}\mathsf{R}-\frac{3}{10}(1+\mathsf{S}) \right] \frac{\Delta \alpha}{\alpha},\\
        \beta_2&=\left[ \frac{4}{5}\mathsf{R}+\frac{17}{10}(1+\mathsf{S}) \right] \frac{\Delta \alpha}{\alpha},\\
        \beta_3&=\left[ \frac{8}{5}\mathsf{R}+\frac{7}{5}(1+\mathsf{S}) \right] \frac{\Delta \alpha}{\alpha}.
    \end{split}
\end{equation}
Here $\beta_1$, $\beta_2$, and $\beta_3$ are the correction terms of $\sqrt{\alpha_\text{p}/\alpha_\text{e}}$,  $\sqrt{\alpha_\text{p}\alpha_\text{e}^3}$, and $\alpha_\text{p}\alpha_\text{e}$, respectively. From the above equations, it is worth noting that $\beta_3=\beta_1+\beta_2$ and thus effectively we are left out only two independent parameters $\beta_1$ and $\beta_2$. In other words, the uncertainties in $\alpha_\text{p}$ and $\alpha_\text{e}$ can be understood through the parameters $\beta_1$ and $\beta_2$.

%%%%%%%%%%%%%%%%%%%%%%%%%%%%%%%%%%%%%%%%%%%%%%%%%%%%%%%%%%%%%%%%%%%%%%%%%%%%%%%%%%%%%%%%%%%%%%%%%%%%%%%%%%%%%%%%%%%%%
\section{Observational constraints on fine-structure constant and proton-to-electron mass ratio}\label{Sec:3}

\begin{figure}[htpb]
    \centering
    \includegraphics[scale=0.5]{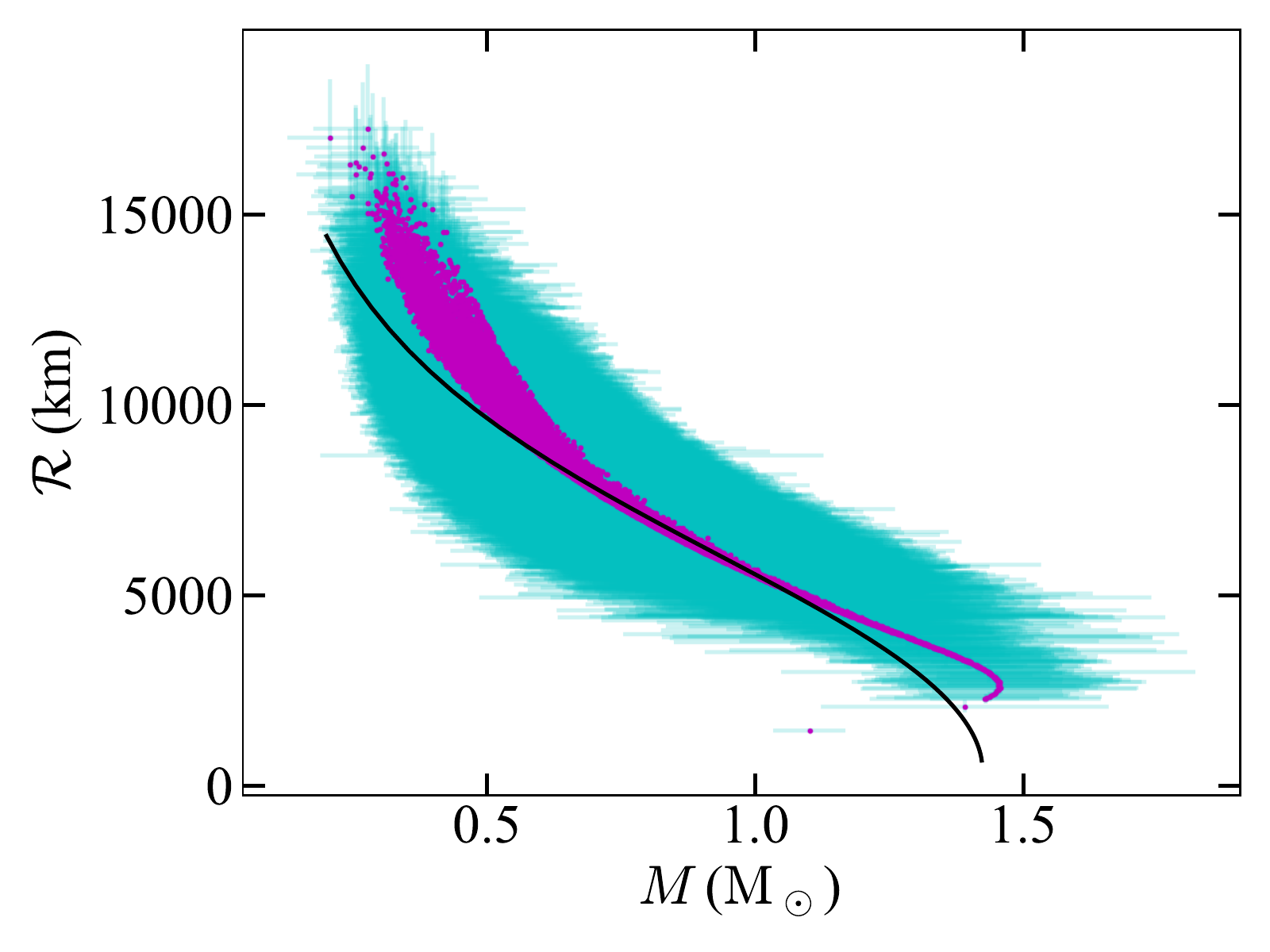}
    \caption{The black line shows the standard Chandrasekhar mass--radius curve for WDs in Newtonian gravity. The magenta scattered points are masses and radii derived from the {\it Gaia}-DR2 observation with their respective error bars shown in cyan color.}
    \label{fig: Gaia}
\end{figure}

In this section, we put constraints on $\alpha$ and $\mu$ using the WDs observed in {\it Gaia}-DR2 survey in the presence of modified gravity. We use masses and radii of the WDs obtained by~\cite{2018MNRAS.480.4505J}. They apparently reported 73\,221 WD candidates using the photometric and astrometric data obtained in this survey. They further used Virtual Observatory Spectral Energy Distribution Analyzer to determine the effective temperatures and luminosities of the WDs and thereby their radii. Further, using the measured $\log\,g$ values with $g$ being the surface gravitational acceleration, they reported masses of these WDs. Fig.~\ref{fig: Gaia} shows masses and radii of these WDs along the Chandrasekhar mass--radius curve for carbon-oxygen core WDs. 
At low density (big-size WDs), there is clearly a deviation from the theoretical curve. This deviation can easily be explained with the inclusion of temperature in the model. One needs to incorporate appropriately the temperature-dependent EoS as well as the temperature gradient equation to obtain the mass--radius curves~\citep{2014PhRvC..89a5801D,2022ApJ...925..133B}. However, as density increases, WDs become smaller in size and the effect of temperature on their masses and sizes also diminishes; thus the Chandrasekhar mass--radius curve is good enough to explain the intermediate-size WDs. Further increase in density results in the deviation from the theoretical curve again and the inclusion of temperature, in general, cannot explain this difference. One can, in principle, include rotation and magnetic fields to explain these WDs. However, from Fig.~\ref{fig: Gaia}, it is clear that these massive WDs (which are over a thousand in number) follow a particular well-defined path, unlike the lighter ones. Thus if the magnetic field or rotation is solely responsible for this phenomenon, all these WDs must have a particular magnetic field strength or a fixed angular speed, which seems unlikely as these WDs are in different positions in the sky. Moreover, from this plot, we observe that for the big and intermediate-size WDs, magnetic fields and rotation do not play a significant role. Furthermore, note that some WDs exceed the Chandrasekhar mass-limit. In this regard, modified gravity theories might be a better bet to explain their structures. It is worth noting that the mass and radius decrease with a further increase in density. This is because, at such a high density, massive elements like nickel, iron, etc. might be present at the core of the WD, which eventually reduces the mass and size of the WDs~\citep{1996cost.book.....G}.

In our previous studies, by obtaining WD mass--radius curves, we showed that modified gravity affects only the massive WDs because of their high core densities~\citep{2018JCAP...09..007K,2021ApJ...909...65K,2022PhRvD.105b4028S,2022PhLB..82736942K,2023PhRvD.107d4072K}. We know that $R$ is nearly proportional to density and thus the $R^2$ term starts showing its effect only at high densities. At low densities, modified gravity does not have any significant effects, and the mass--radius curve overlaps with the standard Chandrasekhar curve. In this work, because our aim is to study the consequences of modified gravity on the constraints of $\alpha$ and $\mu$, we restrict our WD sample to the mass range of $M>\rm M_\odot$ and radius range of $2640\rm\,km<\mathcal{R}<5700\,km$. Now for each WD data, we can theoretically obtain its mass and radius. Thereby we define the following function~\citep{2016PhRvL.116o1103J}
\begin{align}\label{e3.1}
    \chi^2=\sum_{i=1}^N \chi_i^2(M),
\end{align}
where
\begin{equation}\label{e3.2}
    \chi_i^2(M)=\frac{(M-M_i)^2}{\sigma_{M_i}^2}+\frac{(\mathcal{R}_\text{Th}(M)-\mathcal{R}_i)^2}{\sigma_{\mathcal{R}_i}^2},
\end{equation}
with $N$ being the number of data points. Here $M_i$ and $\mathcal{R}_i$ are respectively the mass and radius of $i^\text{th}$ WD in the {\it Gaia}-DR2 data with $\sigma_{M_i}$ and $\sigma_{\mathcal{R}_i}$ being their respective uncertainties. $\mathcal{R}_\text{Th}(M)$ is the theoretical radius corresponding to each $M$ of the WD. We now minimize this $\chi^2$ function for each $\gamma$ value. As $\gamma$ changes, the theoretical mass--radius curve alters and so the $\mathcal{R}_\text{Th}(M)$. In other words, for each $\gamma$, the minimum value of $\chi^2$ is obtained at a particular $\beta_1$ and $\beta_2$. As $\gamma$ changes, $\beta_1$, $\beta_2$, and minimum $\chi^2$ change, which eventually change $\Delta\alpha$.

\begin{figure}[htpb]
    \centering
    \subfigure[$\gamma=0$]{\includegraphics[scale=0.5]{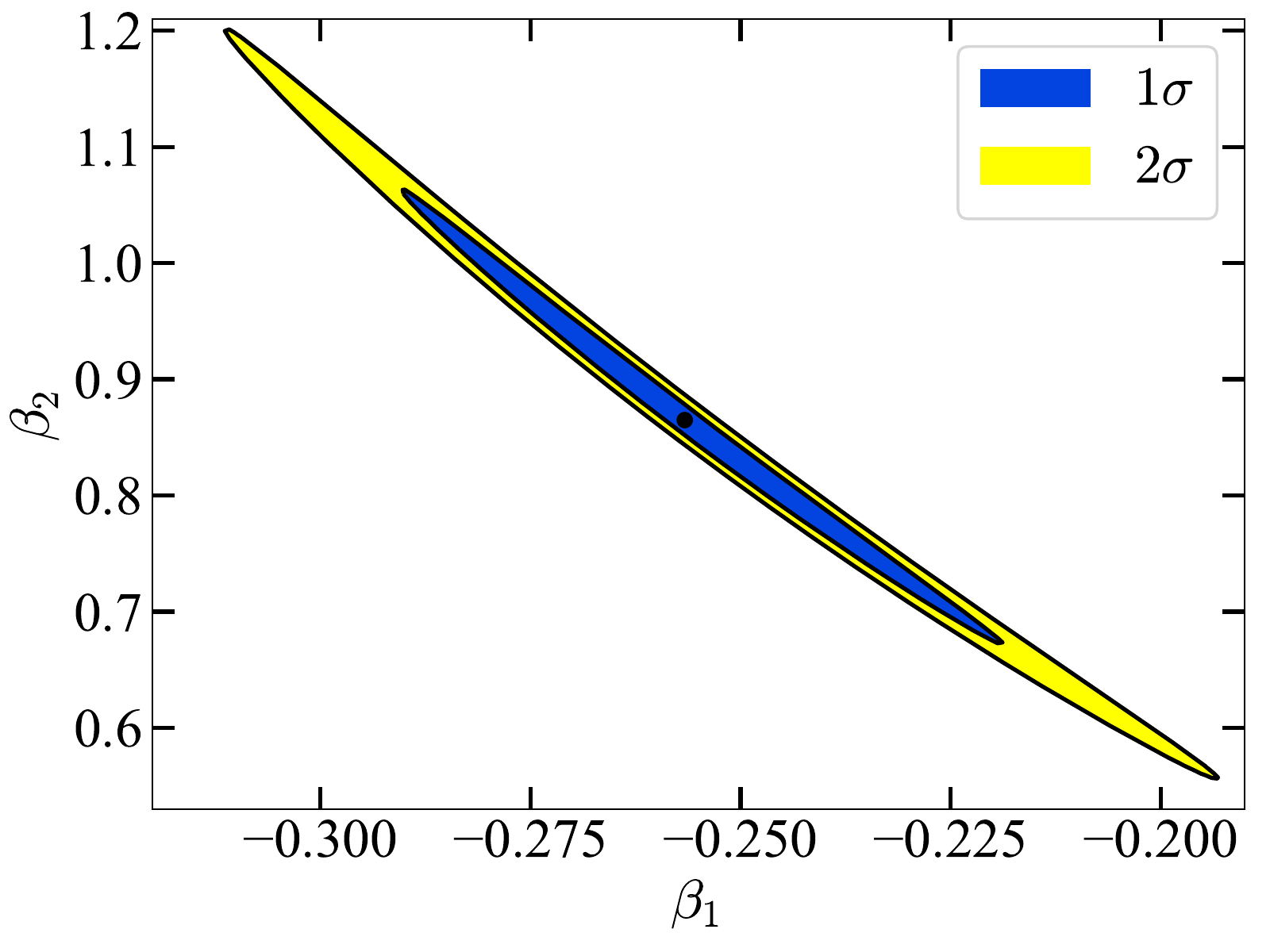}}    
    \subfigure[$\gamma = -1.2\times10^{14}\rm\,cm^2$]{\includegraphics[scale=0.5]{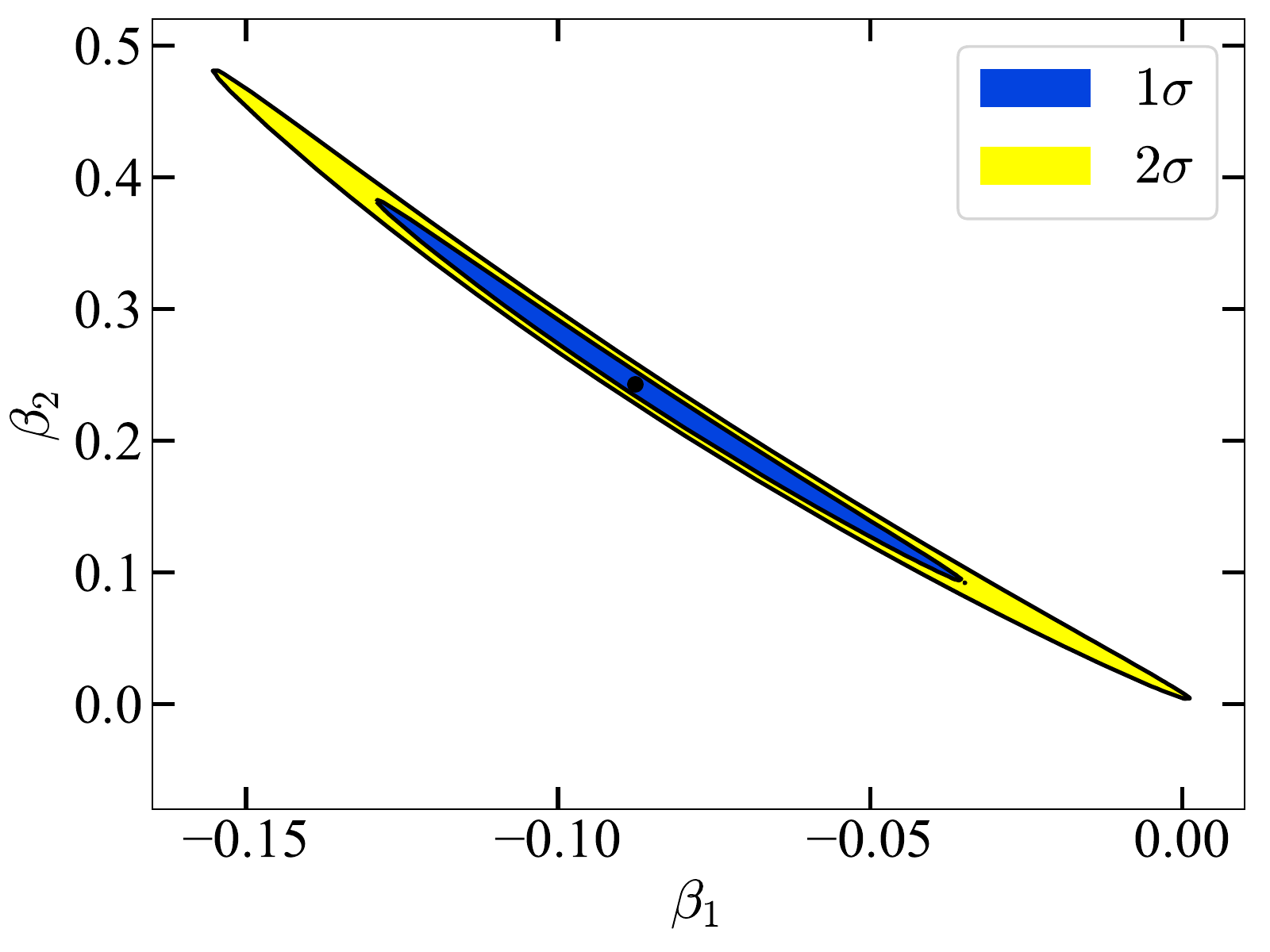}}    
    \caption{Contour plots of $\beta_1$ and $\beta_2$ for $\chi^2$ values. The blue and yellow color respectively shows the $1\sigma$ and $2\sigma$ uncertainty contours and the black dot represents the position for minimum $\chi^2$.}
    \label{fig: contour}
\end{figure}

\begin{figure}[htpb]
    \centering
    \subfigure[$\Delta\alpha/\alpha$]{\includegraphics[scale=0.5]{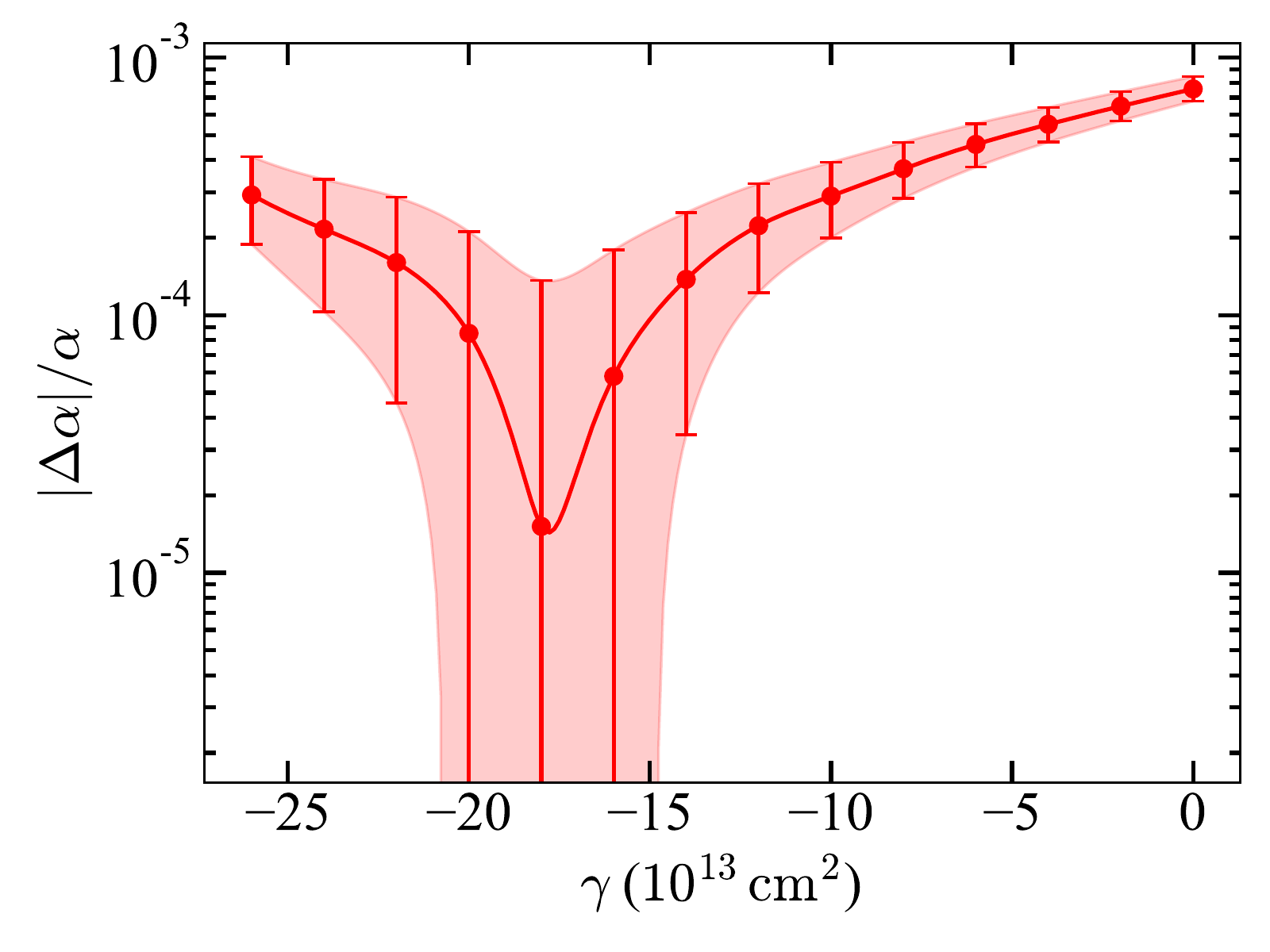}}    
    \subfigure[$\Delta\mu/\mu$]{\includegraphics[scale=0.5]{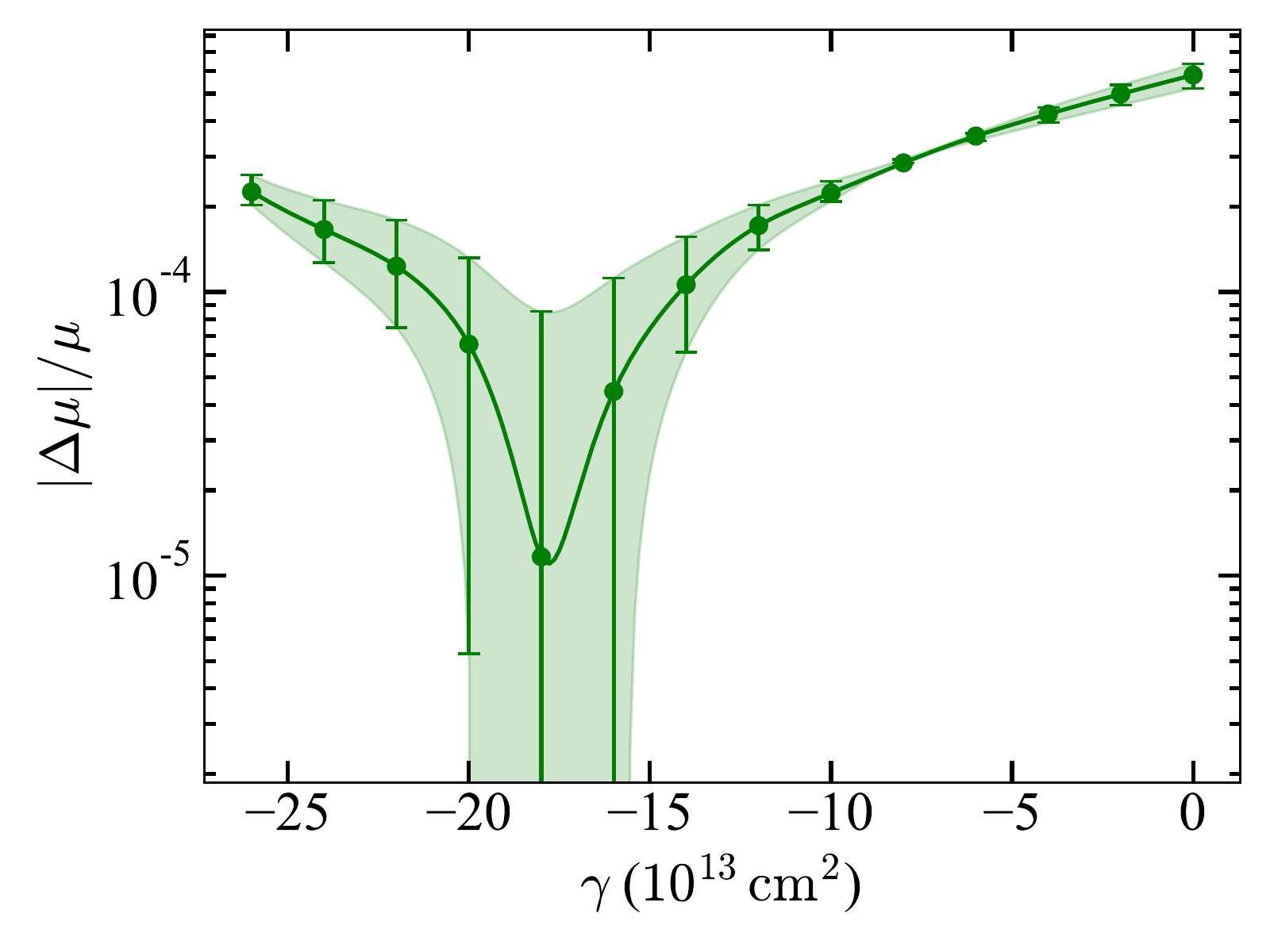}}    
    \caption{Plots of $\abs{\Delta\alpha}/\alpha$ and $\abs{\Delta\mu}/\mu$ along with $1\sigma$ errorbars as a function of $\gamma$.}
    \label{fig: constants}
\end{figure}

Fig.~\ref{fig: contour}(a) shows the contours obtained by simultaneously solving Eqs.~\eqref{e2.17} and~\eqref{e2.18} for $\gamma=0$, i.e. the Newtonian case. We obtain the minimum $\chi^2$ at $\beta_1=-0.256$ and $\beta_2=0.865$, which we use to plot corresponding $1\sigma$ and $2\sigma$ contours. Now, substituting these $\beta_1$ and $\beta_2$ in the first two relations of Eq.~\eqref{e2.19}, we calculate $\Delta\alpha/\alpha$. In this work, we consider $\mathsf{R}=278\pm24$ and $\mathsf{S}=742\pm65$ obtained with $1\sigma$ uncertainty by~\cite{2014MmSAI..85..113M}. Using the errors in $\mathsf{R}$ and $\mathsf{S}$, and extend of the $1\sigma$ contour, we calculate the $1\sigma$ uncertainty for $\Delta\alpha/\alpha$. Further, substituting these values in Eq.~\eqref{e2.13}, we obtain $\Delta\mu/\mu$ and its uncertainty. We repeat the same procedure for different $\gamma$ and for illustration, we show the contours for $\gamma=-1.2\times10^{14}\rm\,cm^2$ in Fig.~\ref{fig: contour}(b). In each case, we obtain $\Delta\alpha/\alpha$ and $\Delta\mu/\mu$ with their respective $1\sigma$ uncertainties. Fig.~\ref{fig: constants} depicts the variations of $\abs{\Delta\alpha}/\alpha$ and $\abs{\Delta\mu}/\mu$ along with their $1\sigma$ errorbars with respect to $\gamma$\footnote{Because we use the logarithmic scale in the plot, some errorbars might look large.}. It is evident that the bounds on these fundamental parameters vary with respect to $\gamma$ implying their dependency on the gravity model.

%%%%%%%%%%%%%%%%%%%%%%%%%%%%%%%%%%%%%%%%%%%%%%%%%%%%%%%%%%%%%%%%%%%%%%%%%%%%%%%%%%%%%%%%%%%%%%%%%%%%%%%%%%%%%%%%%%%%%
\section{Discussion and conclusions}\label{Sec:4}

Variations of fundamental constants like $\alpha$ and $\mu$ can infer the underlying physics of the embedding system. Earlier researchers put forward different bounds for these parameters. Even in the context of modified gravity e.g. $f(T)$ gravity with $T$ being the scalar torsion, the variation of $\alpha$ was already explored~\citep{Wei:2011jw}. Further, using the temporal variation of $\alpha$, the modified gravity parameter in $f(R)$ gravity was constrained~\citep{Bisabr:2010ap}. The majority of these analyses, nevertheless, were inferred from cosmological data and the results alter for different redshifts. Therefore, the energy scale of the system plays a significant role in these bounds. In this exploration, we have used {\it Gaia}-DR2 WD data, which is an astrophysical survey. These observations generally contain information of the surface properties (surface temperature, magnetic fields, etc.) of the compact object and they are not significantly affected by modified gravity due to relatively low densities as compared to that of the core. We have selected only the massive WDs because they are dense and mostly affected by modified gravity. In general, WDs contain magnetic fields and rotation, which can in principle make them heavier. However, to have their significant effects on the WD mass, magnetic-to-gravitational (ME/GE) and kinetic-to-gravitational (KE/GE) energy ratios need to be very high, which may violate the proposed bounds in literature~\citep{1989MNRAS.237..355K,2009MNRAS.397..763B}. Thus even though {\it Gaia}-DR2 WD catalog does not report the magnetic field strength and the rotation speed of these WDs, in case they are affected by modified gravity, unless the ME/GE and KE/GE ratios are significantly high, it would not affect the inferred masses and radii from {\it Gaia}-DR2 survey. As an example, we have further considered $R+\gamma R^2$ gravity model to understand the dependency of bounds for $\alpha$ and $\mu$ on $\gamma$. The robustness of the quadratic Palatini theory was demonstrated by the fact that all charged solutions have a wormhole structure and that these wormhole solutions remain unchanged if one adopts a Palatini approach, which incorporates the spontaneous creation/annihilation of entangled particle–antiparticle pairs~\citep{2014EPJC...74.2924L}. Note that $\gamma$ is an elementary model parameter, and hence as $\gamma$ changes, the gravity theory gets modified. In our case, we have found that the most stringent bounds are $\abs{\Delta\alpha}/\alpha = 1.514^{+12.133}_{-12.304}\times 10^{-5}$ and $\abs{\Delta\mu}/\mu = 1.165^{+7.365}_{-7.471}\times 10^{-5}$ for $\gamma \approx -1.8\times 10^{14}\rm\,cm^2$. This bound is stronger than many of the previously reported bounds mentioned in the Introduction. We have chosen the values of $\gamma$ in such a way that they are within the bounds given by the Gravity Probe B experiment~\citep{2010PhRvD..81j4003N}.

It is worth noting that we do not demand that $R+\gamma R^2$ is the most accurate model of gravity. Rather we have used this model to show the maximum possible variations of $\alpha$ and $\mu$ with respect to $\gamma$. One can, in principle, look for other rigorous modified gravity models to figure out much stronger bounds. However, as mentioned in Sec.~\ref{Sec:2}, our results hold good also for general fourth-order gravity theories and scalar-tensor-vector gravity models. Some other improvements can be made by considering all WDs rather than just choosing the massive ones and then incorporating temperature-dependent models along with the modified gravity, which results in obtaining the temperature-dependent bounds on these parameters under modified gravity. Furthermore, recently {\it Gaia}-DR5 has been released which can be accounted for the improvement of these results. We are developing these ideas and will present the outcomes in the future. To summarize, the bounds of fundamental parameters depend not only on the energy scale but also on the underlying gravity theory.

\begin{acknowledgements}
We thank Aneta Wojnar of Universidad Complutense de Madrid as well as Sayan Chakrabarti and Santabrata Das of IIT Guwahati for useful discussions. SK would like to acknowledge support from the South African Research Chairs Initiative of the Department of Science and Technology and the National Research Foundation as well as the UCT URC Open Access Journal Publication Fund (OAJPF) accredited by the Department of Higher Education and Training (DHET). {\it Computations were performed using facilities provided by the University of Cape Town’s ICTS High Performance Computing team: \href{https://ucthpc.uct.ac.za/}{hpc.uct.ac.za}}.
\end{acknowledgements}


\bibliography{bibliography}{}
\bibliographystyle{aasjournal}



\end{document}